\documentclass[iop]{emulateapj}

\usepackage{epsfig}

\usepackage{graphicx}
\usepackage{float}
\usepackage[caption =false]{subfig}
\usepackage{graphicx}
\usepackage{multirow}

\shorttitle{A Hard Stellar Ionizing Spectrum at z=6.11}
\shortauthors{Mainali et al.}

\begin{document}

\title{Evidence for a Hard Ionizing Spectrum from a z=6.11 Stellar Population}

\email{rmainali@email.arizona.edu}

\author {Ramesh Mainali\altaffilmark{1}, Juna A. Kollmeier\altaffilmark{2}, Daniel P. Stark\altaffilmark{1}, Robert A. Simcoe\altaffilmark{3}, 
Greg Walth\altaffilmark{4}, Andrew B. Newman\altaffilmark{2}, Daniel R. Miller\altaffilmark{3}}

\altaffiltext{1}{Department of Astronomy, Steward Observatory, University of Arizona, 
933 North Cherry Avenue, Rm N204, Tucson, AZ, 85721}
\altaffiltext{2}{Carnegie Observatories, 813  Santa  Barbara  Street,  Pasadena,  CA 91101, USA}
\altaffiltext{3}{MIT-Kavli Center for Astrophysics and Space Research, 77 Massachusetts Avenue, Cambridge, MA 02139, USA}
\altaffiltext{4}{University of California, Center for Astrophysics and Space Sciences, 9500 Gilman Drive, San Diego, CA 92093, USA}

\begin{abstract}
We present the Magellan/FIRE detection of highly-ionized CIV$\lambda$1550 and OIII]$\lambda$1666 in a
deep infrared spectrum of the $z=6.11$ gravitationally lensed low-mass
galaxy RXC J2248.7-4431-ID3, which has previously-known Ly$\alpha$.  No
corresponding emission is detected at the expected location of
HeII$\lambda$1640.  The upper limit on HeII paired with detection of OIII] and
CIV constrains possible ionization scenarios.  Production of CIV and
OIII] requires ionizing photons of 2.5-3.5 Ryd, but once in that state
their multiplet emission is powered by collisional excitation at lower
energies ($\sim 0.5$ Ryd).  As a pure recombination line, HeII
emission is powered by 4 Ryd ionizing photons.  The data therefore
require a spectrum with significant power at $3.5$ Ryd but a rapid
drop toward 4.0 Ryd.  This hard spectrum with a steep drop is
characteristic of low-metallicity stellar populations, and less
consistent with soft AGN excitation, which features more 4 Ryd
photons and hence higher HeII flux.  The conclusions based on ratios
of metal line detections to Helium non-detection are strengthened if
the gas metallicity is low.  RXJ2248-ID3 adds to the growing handful of
reionization-era galaxies with UV emission line ratios distinct from
the general $z=2-3$ population, in a way that suggests hard ionizing
spectra that do not necessarily originate in AGN.
\end{abstract}


\section{Introduction}

Over the past year, the first detailed spectroscopic measurements constraining the nature of 
$z>6$ star forming galaxies have emerged (see \citealt{Stark2016b} for a review), suggesting a different population 
than is common at $z\simeq 2-3$.   Deep near-infrared spectroscopy has revealed  strong UV metal 
line emission in galaxies at $z=6-8$ with equivalent widths 5-10$\times$ larger than are typical at $z\simeq 2$ 
\citep{Stark2015a,Stark2015b,Stark2016}, while ALMA observations have begun to deliver detections of [CII] 158$\mu$m and [OIII] 88$\mu$m
emission in typical galaxies at $z>6$ \citep{Willott2015, Pentericci2016, Inoue2016, Bradac2016}.
Perhaps the most surprising result is the discovery of nebular CIV$\lambda\lambda1548,1550$ in a galaxy at $z=7.045$ 
 \citep{Stark2015b}, requiring an extremely hard radiation field capable of producing a large number of photons 
 more energetic than 47.9 eV.    Only one percent of UV selected galaxies at $z\simeq3$ show strong nebular 
CIV emission \citep{Steidel2002, Reddy2008,Hainline2011}.  While these systems 
tend to be low luminosity narrow line AGNs, more recent studies have shown that nebular CIV is also found 
in dwarf star forming galaxies \citep{Christensen2012, Stark2014, Vanzella2016}, 
presumably powered by the harder radiation field from low metallicity stars.  
The detection of nebular CIV  in 
one of the first  galaxies targeted  in the reionization era suggests that galaxies with hard ionizing spectra may 
be  more common in the reionization era. 

\begin{figure*}
\centering
\subfloat{\includegraphics[scale=0.25]{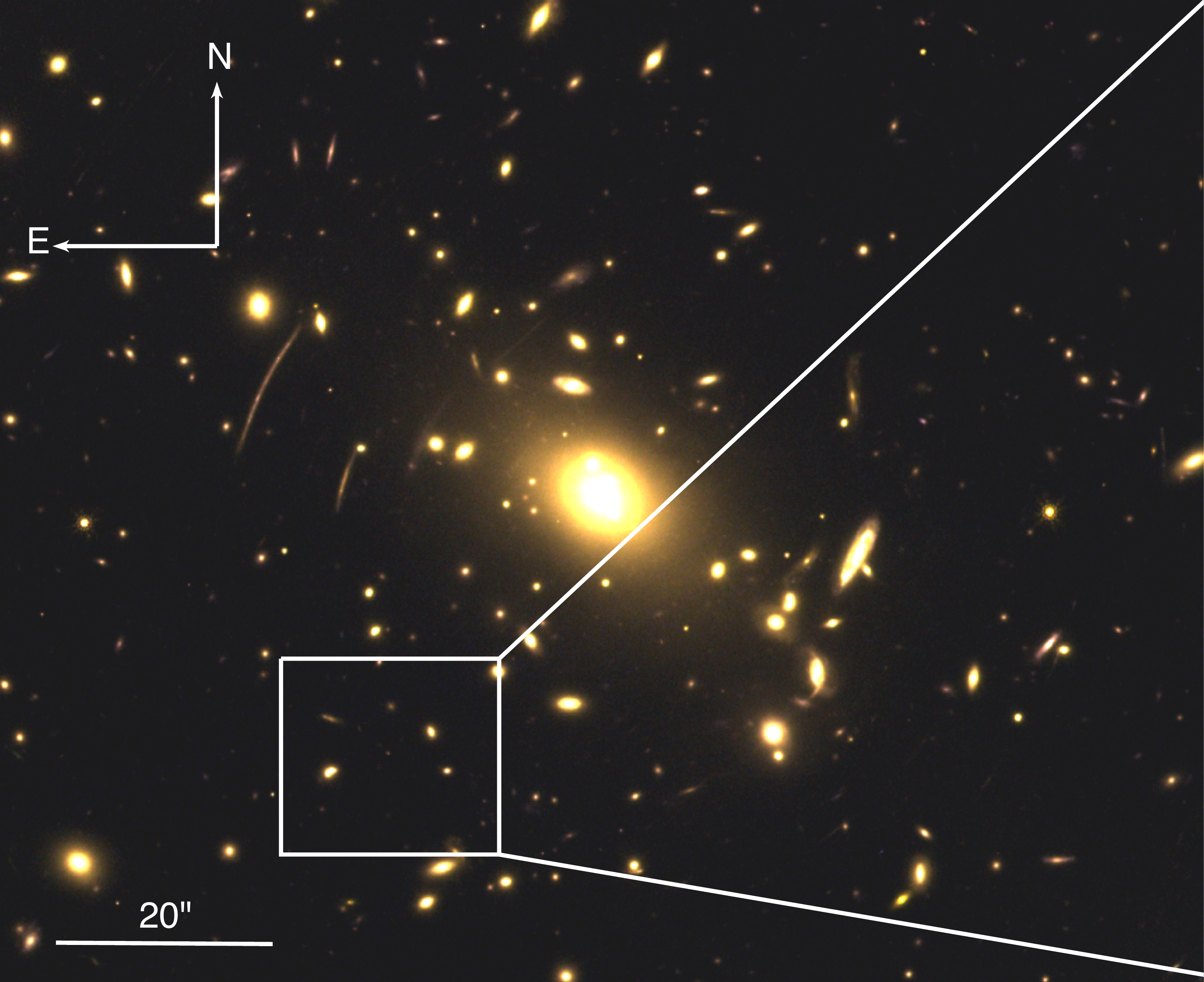}}\
\subfloat{\includegraphics[scale=0.246]{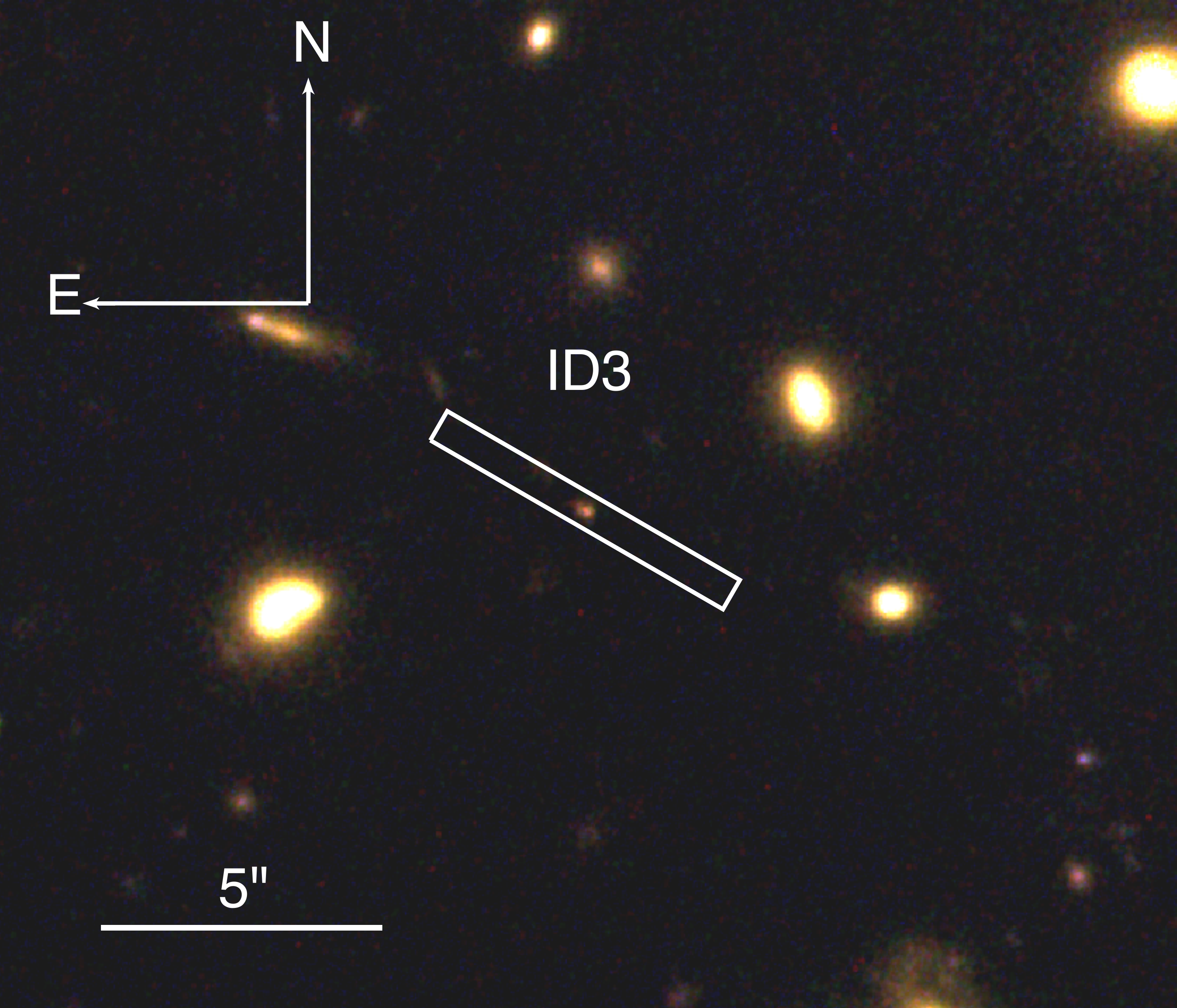}}\\

\caption{(Left:) {\it HST} WFC3/IR  color image of the cluster RXC J2248 showing  
the position of the galaxy RXC J2248 - ID3. (Right:)  Slit center and position 
angle of the Magellan/FIRE observations.}
\end{figure*}

There are two outstanding issues that must be addressed  following these preliminary spectroscopic studies.  First, it remains unclear 
how  representative  the $z=7.045$ CIV emitter is of star forming galaxies at $z>7$.  
If stellar populations commonly produce hard ionizing spectra at
$z>6$, it would represent a rapid and qualitative change in the galaxy
population relative to all lower redshifts, and these galaxies would
play a larger role in reionization than has previously been assumed.
Second, one must attempt to establish whether the high-ionization
emission is powered by hot, low-metallicity stars or AGN. Both can
potentially provide high energy photons, and with a single metal-line
detection it is difficult to prove the source beyond a reasonable
doubt (e.g. Stark et al 2015b).
  At lower redshifts, the separation of AGN and star forming galaxies is readily carried out 
using rest-frame optical emission line ratios (e.g., \citealt{Baldwin1981}). However utilization of a 
similar approach at $z>6$ must await  the launch of JWST.   Recent 
efforts have begun to develop rest UV diagnostics based on different photoionization models to distinguish 
the sources of ionizing spectrum \citep{Feltre2016}.  This can be achieved with current ground-based facilities, 
provided multiple far-UV lines can be detected.  

In this paper, we describe initial results from a  spectroscopic campaign using the Magellan Baade Folded-port InfraRed Echellette 
(FIRE; \citealt{Simcoe2013}) aimed at addressing the two issues described above.  The spectral coverage provided by FIRE 
makes it particularly efficient at recovering multiple lines in bright $z>6$ galaxies.      Here we describe  FIRE 
observations of a $z=6.110$ gravitationally-lensed galaxy.   The FIRE spectrum reveals the presence of the nebular CIV$\lambda$1550 emission line, providing 
another instance of high ionization lines at  $z>$ 6.   We also report  
detection of a second feature (OIII]$\lambda\lambda$1660,1666), enabling exploration of the origin of the high 
ionization emission.

Throughout the paper we adopt standard $\Lambda$CDM cosmology 
with $\Omega_{M}$=0.3, $\Omega_{\Lambda}$=0.7, H$_{0}$=100 h km s$^{-1}$ Mpc$^{-1}$, and h=0.7. 
Magnitudes are quoted in AB magnitudes. 

\section{Observations}

\begin{table}
\begin{tabular}{lcccc}
\hline 
 & $\rm{\lambda_{rest}}$$^{a}$  &$\rm{\lambda_{obs}}$  &  Line Flux & $\rm{W_{0}}$ \\ 
 & ($\rm{\AA}$) & ($\rm{\AA}$) & ($\rm{10^{-18}erg~ cm^{-2} ~ s^{-1}}$) & ($\rm{\AA}$) \\ \hline 
 \multicolumn{5}{c}{Magellan/FIRE} \\\hline 
Ly$\rm\alpha$ & 1215.67 & 8643.5 & 33.2$\pm$2.3 & 39.6$\pm$5.1 \\
 NV & 1240 & \ldots & $<1.8$ & $<2.3$\\
CIV  & 1548.19 & \ldots & \ldots & \ldots \\
\ldots &1550.77 & 11023.8 & 5.7$\pm$0.9 & 9.9$\pm$2.3 \\
He II & 1640.42 & \ldots & $<1.5$ & $<2.8$ \\
OIII] & 1660.81 & 11796.9 & 1.7$\pm$0.6 & 2.9$\pm$1.4\\
\ldots & 1666.15 & 11837.1 & 2.7$\pm$0.6 &4.6$\pm$1.6 \\ \hline 
 \multicolumn{5}{c}{ \textit{HST} NIR WFC3 G102/G141} \\ \hline  

 CIV & 1549$^{b}$ & \ldots & 14.0$\pm$3.8 & 24.5$\pm$7.1\\
 OIII] & 1663$^{c}$ & \ldots & $<7.6$ & $<13.0$\\
  CIII] & 1908$^{d}$ & \ldots & $<3.6$ & $<7.9$\\
  
\hline\hline
\end{tabular}
\\
$^{a}$vacuum wavelengths, $^{b}$Total CIV$\lambda\lambda$1548,1550 flux,  $^{c}$Total OIII]$\lambda\lambda$1660,1666 flux,
$^{d}$Total CIII]$\lambda\lambda$1907,1909 flux.

\caption{Rest-UV emission line measurements of the $z=6.11$ galaxy RXC J2248-ID3 from Magellan/FIRE and the HST WFC3/IR grism.  
Non-detections are listed as 2$\sigma$ upper limits. }
\label{table:nir}
\end{table}

\begin{figure*}
\centering
\subfloat{\includegraphics[scale=0.72]{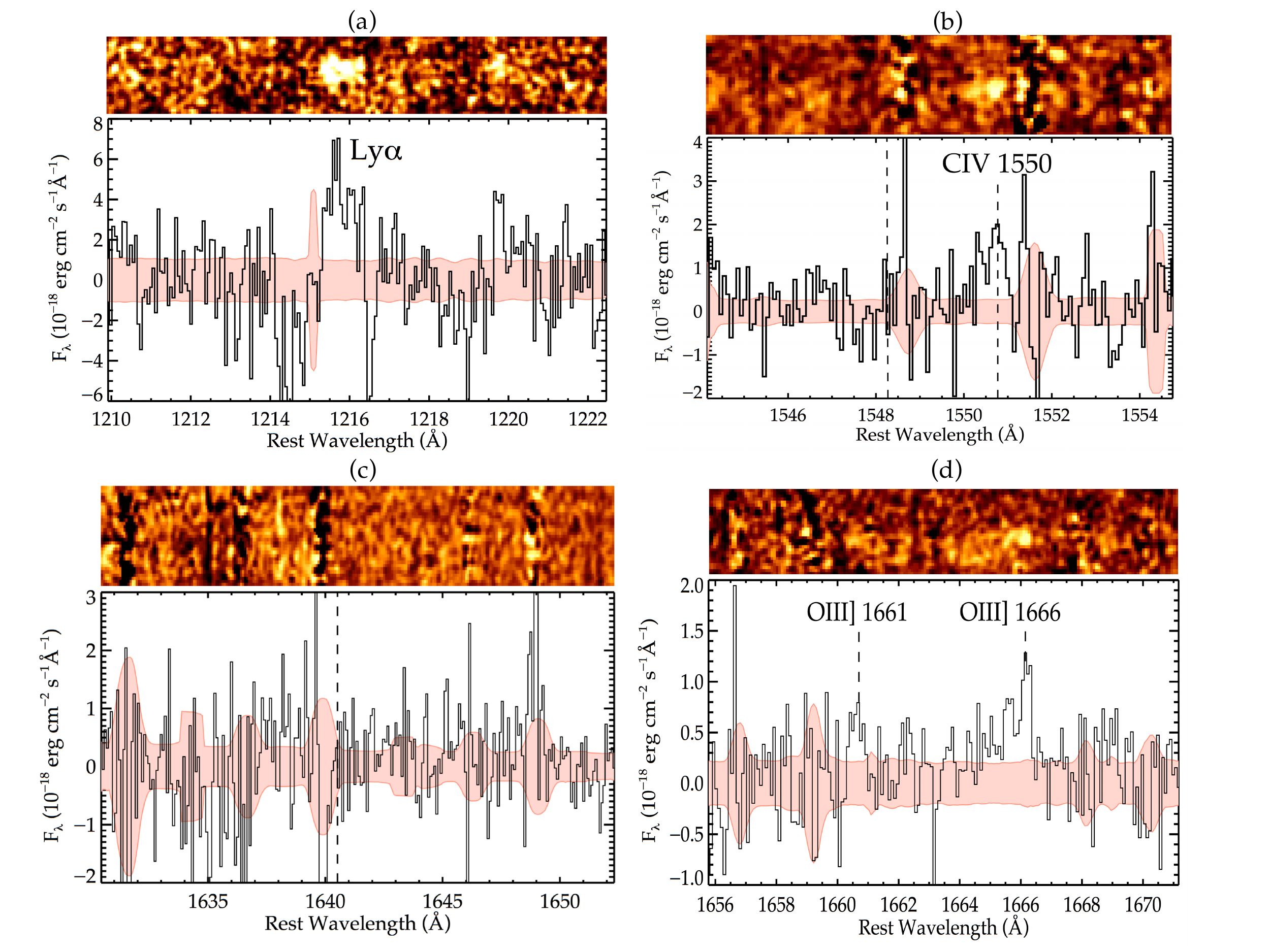}}\\
\subfloat{\includegraphics[scale=0.39]{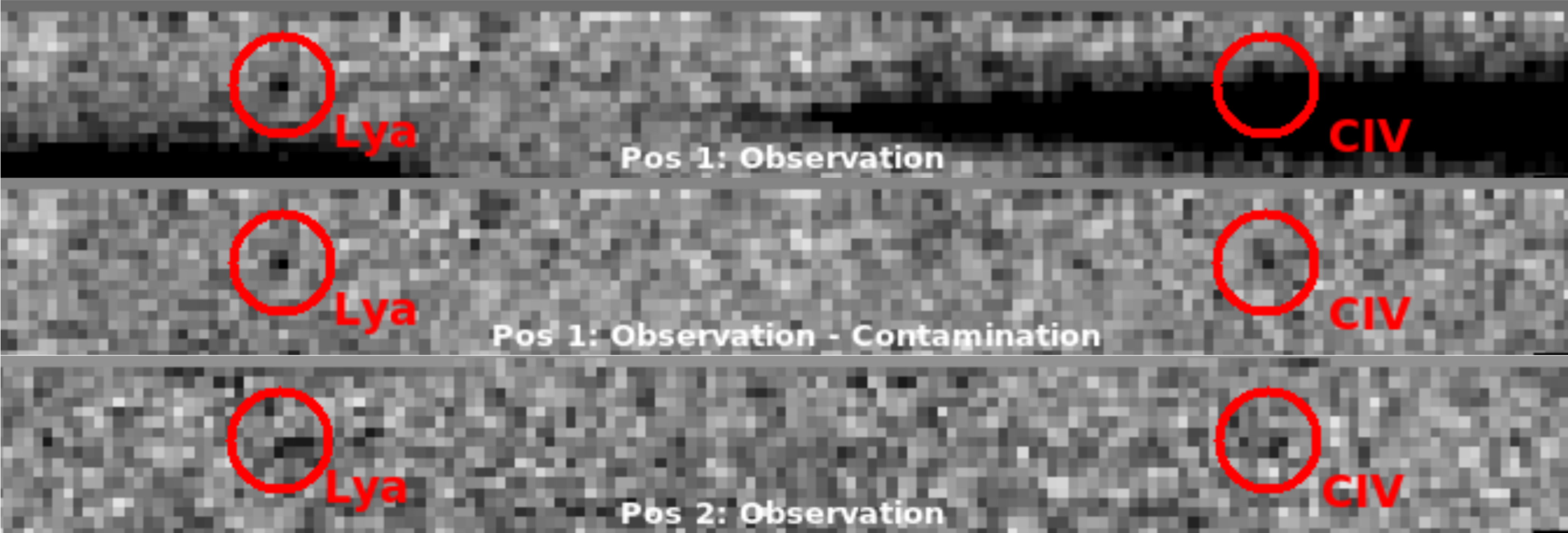}}\\

\caption{(Top:) Magellan/FIRE spectrum (R=6000) of the $z=6.11$ galaxy RXC J2248-ID3.  Each panel shows the 2D spectrum (with white corresponding to positive flux) on top of the 1D spectrum.  The error in the 1D spectrum is shown in red.  
 (Bottom:) \textit{HST} WFC3/IR G102 spectra (R=210) of RXC J2248-ID3 at two different roll angles (black corresponds to positive flux), with the middle panel showing the spectrum following contamination subtraction.  }
\end{figure*}

We report on observations of RXC J2248-ID3, one of five images of a $z=6.11$ gravitationally lensed galaxy behind the cluster 
RXC J2248.7-4431.  The galaxy was first identified by \citet{Boone2013} and \citet{Monna2014} via the Cluster 
Lensing And Supernova survey with Hubble (CLASH, \citealt{Postman2012}).
 The images are bright (J$_{\rm{125}}$=24.8-25.9) owing to their magnification (2.2-8.3$\times$).  
We calculate the UV continuum slope using  deeper imaging from the  
Hubble Frontier Field initiative \citep{Lotz2014}.   The data show 
that the galaxy is very blue  ($\beta$=-2.54$\pm$0.16) and that the  UV absolute magnitude  is $M_{\rm{UV}}=-20.1\pm0.2$ after magnification correction, indicating a sub-L$^\star$ luminosity. As in our previous studies (i.e., \citealt{Stark2013a, Stark2016}), we perform SED fitting using a code developed in \citet{Robertson2010} which combines  \citet{Bruzual2003} with nebular 
line and continuum emission computed assuming case B recombination and  empirical metallic line intensities from \citet{Anders2003}.   
 For a Salpeter stellar IMF and constant star formation, the magnification-corrected data suggest  a low  stellar mass 
 ($1\times10^{8}$ M$_{\odot}$),  little  reddening 
 (E(B-V)=0.01), and a large sSFR (50 Gyr$^{-1}$).   
 Strong Ly$\alpha$ (W$_{\rm{Ly\alpha}}=$ 79~\AA) has been identified in RXC J2248-ID3 by VLT/VIMOS \citep{Balestra2013}, VLT/FORS \citep{Boone2013}, and the {\it HST} WFC3/IR grism \citep{Schmidt2016}, consistent with expectations for young, metal poor galaxies (e.g., \citealt{Cowie2011, Trainor2016}).

UV metal lines tend to be most prominent in systems with extremely large 
EW Ly$\alpha$ emission (e.g., \citealt{Shapley2003, Stark2014}), making RXC J2248-ID3 an ideal candidate for detecting 
metal lines at $z>6$.   At the redshift of  RXC J2248-ID3, the FIRE spectrum is able to constrain the relative strengths of the 
Ly$\alpha$, CIV$\lambda\lambda$1548,1550, He II$\lambda$1640 and OIII]$\lambda\lambda$1660,1666 lines.  Unfortunately, the CIII]$\lambda\lambda$1907,1909 doublet is situated in a region of low atmospheric transmission between the 
J and H-band, precluding useful flux constraints. 

We observed RXC J2248-ID3 over July 19-21 2014 using FIRE in echelle mode, providing  continuous 
spectral coverage between 0.82 and 2.51 $\mu$m. 
The particular image we observed  is magnified by 5.5$\times$.
We adopted a slit width of 0\farcs6, resulting in a resolving power of R=6000. The orientation of the slit  on the galaxy (PA=60$^{\circ}$) is shown 
in Figure 1. The exposures were carried out using two dither positions separated by 3\farcs0.  Observing conditions  were 
excellent with clear sky and an average seeing of 0\farcs4.  Given the seeing, source size, and slit width, we require a 
small aperture correction (1.1$\times$) to the observed fluxes.
The total on-source integration time over three nights was 9.17 hours. 

The FIRE spectrum was reduced using standard routines in the FIREHOSE data reduction 
pipeline\footnote{wikis.mit.edu/confluence/display/FIRE/FIRE+Data+Reduction}. The pipeline uses lamp and sky flats. Two dimensional sky models are iteratively calculated following \citet{Kelson2003}. The wavelength solutions are provided by fitting OH skylines 
in the spectra. Flux calibration and telluric corrections to the data are applied using  A0V star observations.  Finally the 
1D spectra were extracted by using a boxcar  with aperture of 1\farcs35 (15 pixels), corresponding 
to the spatial extent of the strongest emission line (Ly$\alpha$) in the FIRE spectrum. 

The {\it HST} grism spectra for RXC J2248 comes from the  Grism
Lens-Amplified Survey from Space (GLASS, Schmidt et al. 2014; Treu et al. 2015)
survey. The {\it HST} WFC3/IR grisms G102 and G141 have a spectral
resolution of 210 and 130 and cover the wavelengths 0.8-1.15 $\mu$m and
1.1-1.7 $\mu$m. The data were reduced
using aXe (K{\"u}mmel et al. 2009). MultiDrizzle was used to combine the
direct images. Multiple visits at similar roll angles were drizzled together and
tweakshifts was used to determine the offset between the visits. Next,
SExtractor was run on the direct images to generate aXe input catalog. Finally,
the aXe routines are run to drizzle the 2D spectra and extract the spectra. 
The 1D spectra were extracted
from the 2D spectra using a 0.38\arcsec (3 pixels) aperture.  The extracted 1D
spectra are then divided by the instrument sensitivity function and pixel size
to flux calibrate the spectra.  The first roll angle (97.7$^\circ$) had significant
contaminating continuum; a sliding median with a window of 50 pixels (1200 \AA)
was used to subtract off the continuum source.

\begin{figure*}
\centering

\subfloat{\includegraphics[scale=0.32]{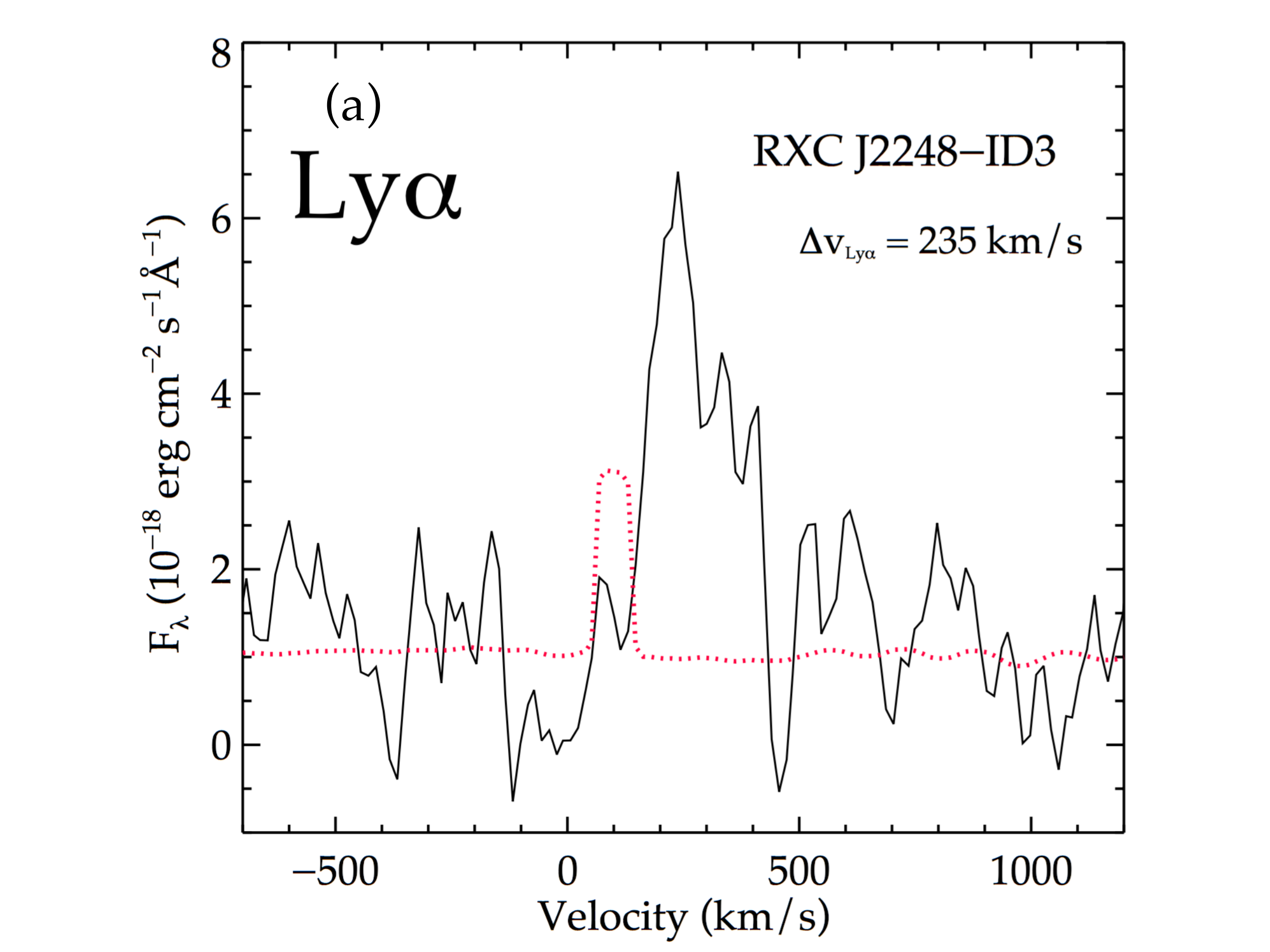}}\
\subfloat{\includegraphics[scale=0.32]{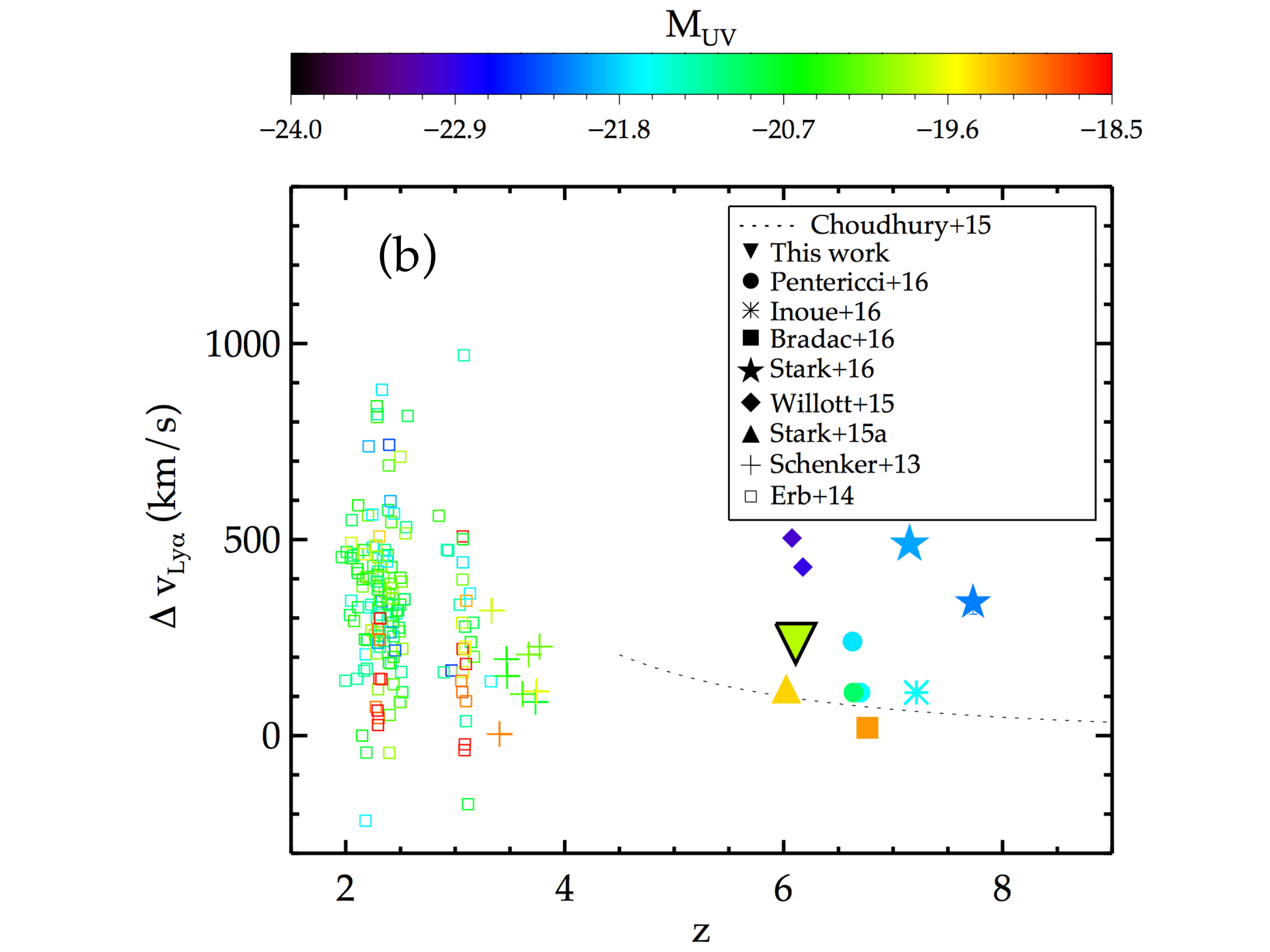}}\\
\caption{(Left:) Lyman $\alpha$ velocity profile of RXC J2248-ID3 derived using the systemic redshift from OIII]$\lambda$1666.
(Right:) Ly$\alpha$ offset velocity ($\rm\Delta v_{Ly\alpha}$) as a function of redshift. The dotted line represents velocity offset model used in \citet{Choudhury2015}}.
\end{figure*}

\section{Results} 

We summarize line measurements  in Table 1.  Strong Ly$\alpha$ emission is clearly visible at 8643.5 $\rm\AA$ in the FIRE spectrum (Figure 2a), implying a redshift 
($z_{\rm{Ly\alpha}}=6.110$) that is consistent with the Ly$\alpha$ wavelength presented in \citet{Balestra2013}.   Using 
the redshift and spatial position defined by Ly$\alpha$, we search for and characterize the strength of other emission lines 
in the FIRE spectrum.   The nebular CIV$\lambda\lambda$1548,1550 doublet is a resonant feature and is often 
scattered away from line center.    Observations of metal-poor star forming galaxies with strong UV metal lines 
show that the peak CIV flux often has the same velocity offset from systemic as Lyman $\alpha$ \citep{Stark2015b, Vanzella2016}.   
For a 10~\AA\ window centered at the Ly$\alpha$ redshift,  CIV$\lambda$1548 line would be located  at 
11003 $\rm\AA$ and 11013 $\rm\AA$ and CIV$\lambda$1550 would lie between 11021 $\rm\AA$ and 11031 $\rm\AA$.   
We clearly detect the  CIV$\lambda$1550 component  with S/N=6.3 in the expected window at 11023.8 $\rm\AA$ (Figure 2b). The line is  barely resolved with FWHM (corrected for instrumental resolution) of 59 km s$^{-1}$.
The total integrated flux  is  5.7$\pm$0.9$\times$$\rm10^{-18}$ erg cm$^{-2}$ s$^{-1}$.   As is standard practice for 
faint high redshift spectra lacking continuum detections, we compute the underlying continuum flux density at 1550~\AA\ using 
the SED model that provides the best fit to the  photometry (see \S2 for details). This translates to a rest frame equivalent width of 9.9$\pm$2.3 $\rm\AA$, where the error includes the uncertainty in the continuum and line measurements. As is evident  in Figure 2b, 
a strong OH skyline is present at the expected location of  CIV$\lambda$1548, precluding useful flux 
constraints for the blue component of the doublet. Observations of metal poor CIV emitters often show equal flux in 
the two components of the doublet (e.g., \citealt{Christensen2012,Stark2014}).  If this is also the case for RXC J2248-ID3, 
we would expect a total CIV line flux of 11.4$\times$10$^{-18}$ erg cm$^{-2}$ s$^{-1}$.  The theoretically-expected flux ratio 
(CIV$\lambda$1548/CIV$\lambda$1550=2) has also been observed in several  intermediate-redshift systems (e.g., \citealt{Vanzella2016,Caminha2016}) and would instead predict a total flux of 
17.1$\times$10$^{-18}$ erg cm$^{-2}$ s$^{-1}$.   
 
Unlike the resonant CIV line, the nebular He II$\lambda$1640 and OIII]$\lambda\lambda$1660,1666 lines typically 
trace the systemic redshift \citep{Shapley2003, Steidel2010, Stark2014}.   Assuming Lyman 
$\alpha$ velocity offsets ($\rm\Delta v_{Ly\alpha}$) between 0 km s$^{-1}$ and 450 km s$^{-1}$, consistent with previous 
studies \citep{Tapken2007, Erb2010,Stark2015b, Stark2016}, we predict that OIII]$\lambda$1660 will fall between 
11794 $\rm\AA$ and 11808 $\rm\AA$ and the OIII]$\lambda$1666  will lie between 11832 $\rm\AA$  
and 11846 $\rm\AA$.   We detect a 4.5$\sigma$ emission  feature (FWHM=58 km s$^{-1}$) 
centered at 11837.1 $\rm\AA$ (see Figure 2d) with total  flux 
2.7$\pm$0.6$\times$$\rm10^{-18}erg~cm^{-2}~s^{-1}$.  We identify this line as OIII]$\lambda$1666, indicating a 
systemic redshift of $z=6.1045$ for RXC J2248-ID3.   Following the same procedure described above for CIV, we 
calculate a rest-frame equivalent width of  4.6$\pm$1.6 $\rm\AA$.  Using the OIII]$\lambda$1666 redshift, we search 
for emission associated with OIII]$\lambda$1660.   A  faint emission component (S/N=2.8)  is visible at the expected 
wavelength (Figure 2d).   The line flux and rest-frame equivalent width are 1.7$\pm$0.6$\times$$\rm10^{-18}erg~cm^{-2}~s^{-1}$ 
and 2.9$\pm$1.4 $\rm\AA$, respectively. Based on the presence of CIV, we may also expect to see nebular He II.   
While the line is expected to  fall in a clean region of the FIRE spectrum at 11655~\AA\ (based on the systemic redshift), there 
is no evidence of any emission feature at the expected location (Figure 2c), implying a 2-$\sigma$ upper limit on the rest frame equivalent width 
of 2.8 $\rm\AA$. The non detection of He II is consistent with the upper limits ($<$1.4 $\rm\AA$) derived for young metal poor galaxies 
at $z\sim2-3$ (e.g., \citealt{Stark2014}).   Similarly we do not detect the NV$\lambda\lambda$1238,1240 line  (W$_{\rm{NV}}<$2.3 $\rm\AA$) 
that is commonly seen in AGN spectra.  

 The systemic redshift provided by the detection of OIII] allows  Ly$\alpha$  to be shifted to the galaxy rest-frame.  Figure 3a 
 shows the resultant Ly$\alpha$ line profile.   The 
 peak Ly$\alpha$ flux is redshifted from systemic by a velocity offset of  $\rm\Delta v_{Ly\alpha}$=235 km s$^{-1}$.  
  The FWHM of the Ly$\alpha$ (131 km s$^{-1}$) is narrower than many luminous reionization-era galaxies with Ly$\alpha$ detections (e.g., \citealt{Oesch2015}).  
 The difference with respect to the CIV and OIII] FWHM is not surprising given the reprocessing of the line profile by the CGM and surrounding IGM.
 Including RXC J2248-ID3, there are now  11 $z>6$ galaxies with Ly$\alpha$ profile and velocity offset measurements (see Figure 3b) where either [CII] 158$\mu$m or [OIII] 88$\mu$m~\citep{Willott2015, Pentericci2016, Inoue2016, Bradac2016} or UV metal line detections \citep{Stark2015a, Stark2016} 
 constrain the systemic redshift.   We discuss trends with M$_{\rm{UV}}$ and implications for 
 the escape of Ly$\alpha$ from  reionization era galaxies in \S4.  
 
  The WFC3/IR grism spectrum of RXC J2248-ID3 is shown along the bottom panel of Figure 2.    Ly$\alpha$  is clearly detected, as was 
 previously reported in \citet{Schmidt2016}.   The spectrum also reveals detection of nebular CIV (unresolved) in both roll angles.  
  The mean integrated flux (14.0$\pm$3.8$\times$$\rm10^{-18}~erg~cm^{-2}~s^{-1}$) and equivalent 
 width (24.5$\pm$7.1 $\rm\AA$)  of the two roll angles  thus reflects both  CIV$\lambda$1548 and CIV$\lambda$1550.  
 The grism spectra  covers He II, OIII], and CIII], but no detections are apparent.  The CIV to CIII] flux ratio ($>$3.9 at 2$\sigma$) is slightly larger 
than the range (0.4-1.6) spanned by metal poor CIV emitters at moderate redshifts \citep{Stark2014,Vanzella2016} but is consistent with 
flux ratios expected for galaxies powered by low metallicity stars \citep{Feltre2016}.
The total CIV flux is  consistent with the flux range (11.4-17.1$\times$$\rm10^{-18}~erg~cm^{-2}~s^{-1}$) predicted from the  detection of the single CIV$\lambda$1550 component FIRE spectrum.  Since the FIRE constraints on the OIII], CIV, and He II fluxes are made under the same atmospheric conditions and are 
subject to the same aperture corrections, we will limit our investigation to line ratios calculated from FIRE.  In the empirically motivated 
case where CIV$\lambda$1548 / CIV$\lambda$1550=1, we would expect log(He II/CIV)$<-0.87$ and log(OIII]/CIV)$=-0.39$.
The grism measurement places an upper bound on the total CIV flux (owing to slit losses), suggesting log(He II/CIV)$<-0.96$ and  log(OIII]/CIV)$=-0.51$.    
We will consider both options in the following section.
  
\section{Discussion and summary}

\begin{figure*}
\centering
\subfloat{\includegraphics[scale=0.32]{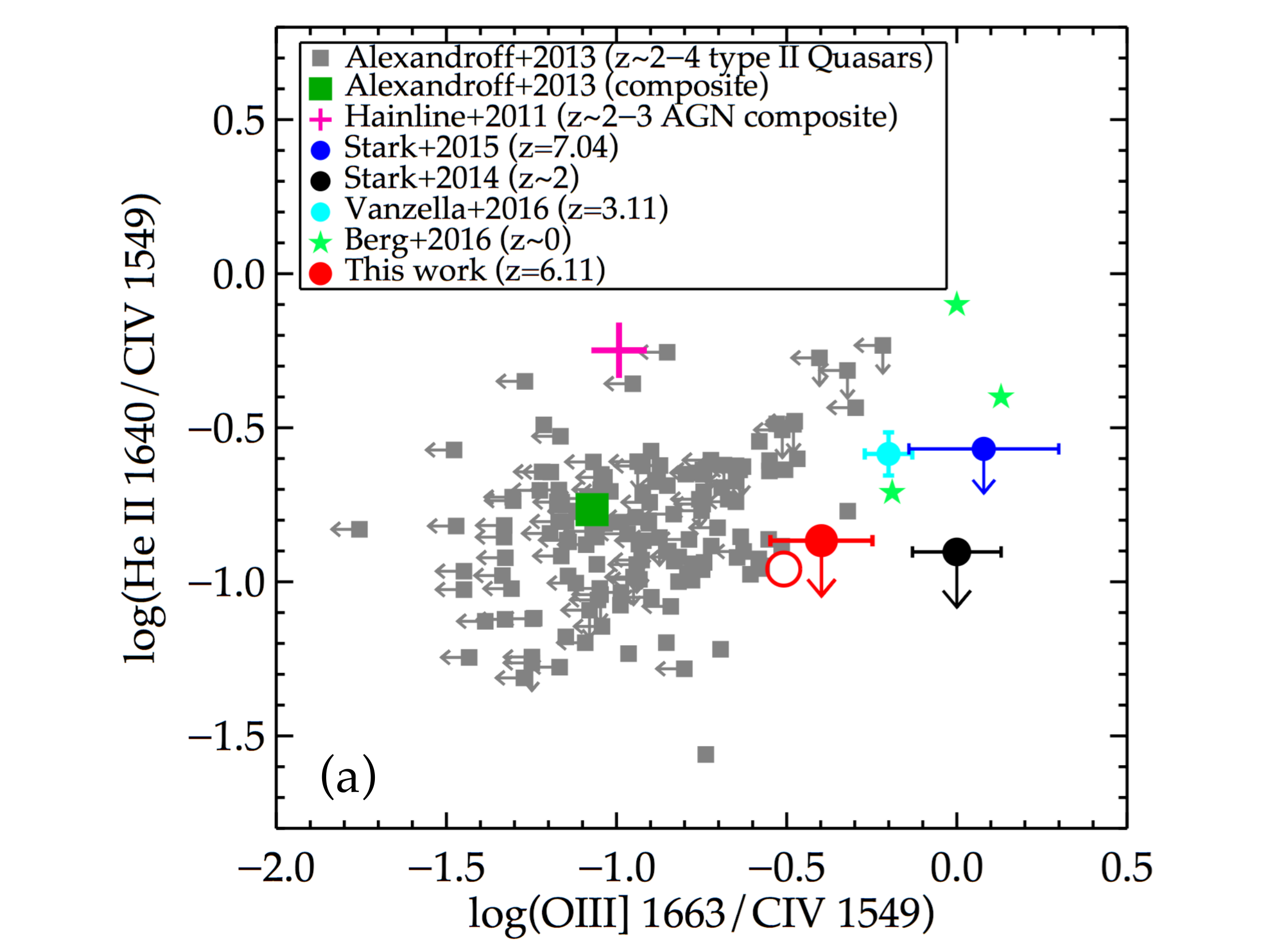}}\
\subfloat{\includegraphics[scale=0.32]{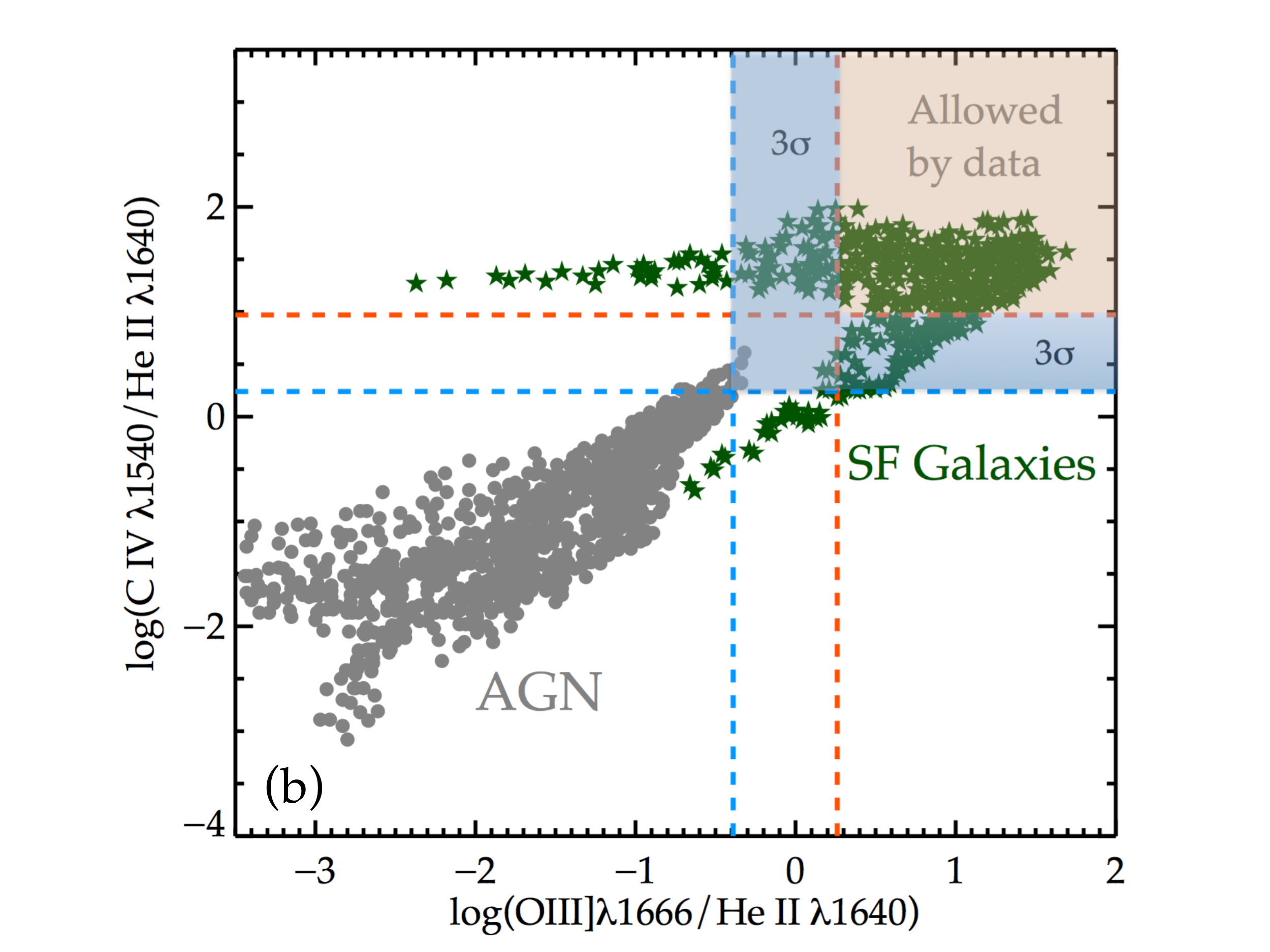}}\\
\caption{(Left:) Comparison of UV line ratios associated with metal poor CIV emitters  narrow line AGN  at z$\sim$2-4.  The filled 
red circle shows the line ratios of  RXC J2248-ID3 if we adopt the empirically-motivated line ratio (CIV$\lambda$1548/CIV$\lambda$1550=1); the 
open red circle corresponds to an upper bound on the total CIV$\lambda\lambda$1548,1550 flux adopted using the WFC3/IR grism measurement.  
The metal poor star forming systems are mostly separated from the AGN samples, suggesting they are subject to a softer radiation field. (Right:) Comparison to photoionization models of \citet{Feltre2016}. Grey(green) points correspond to flux ratios predicted from the AGN(stellar) photoionization 
models in  \citet{Feltre2016}. The red (blue) dashed line represents 1-$\sigma$ (3-$\sigma$) lower limits on the line ratios, 
demonstrating that the data are better explained by a stellar radiation field.}
\end{figure*}

The discovery of nebular CIV emission in RXC J2248-ID3 provides the second 
example of high ionization lines associated with an intrinsically faint lensed reionization-era galaxy.  
Yet the origin of the UV emission features remains poorly understood.  
The presence of multiple lines in the spectrum RXC J2248-ID3 provides a unique opportunity to 
examine the powering mechanism of the high ionization emission.  
Production of nebular CIV and OIII] emission requires ionizing photons in the range 2.5-3.5 Ryd, 
 but once in that state their multiplet emission is powered by collisional excitation at lower
energies ($\sim 0.5$ Ryd).   As a pure recombination line, HeII$\lambda$1640
emission is powered by 4 Ryd ionizing photons.   Photons with 4 Ryd are also capable of triply ionizing  
oxygen, thereby decreasing the strength of OIII] emission.

The presence of strong OIII] and CIV emission  indicates that RXC J2248-ID3  must 
have an ionizing spectrum with significant power at $2.5-3.5$ Ryd.  But a rapid  drop toward 4.0 Ryd is required 
to maintain strong OIII] emission and explain the non detection of He II line.  Such a spectral break is inconsistent with a  
shallow AGN power law spectrum.   The left panel of Figure 4 demonstrates this empirically.
RXC J2248-ID3 and other metal poor CIV emitters \citep{Stark2014,Vanzella2016,Stark2015b,Berg2016}
 have larger OIII]/CIV flux ratios than both 
$z\simeq 2-3$ UV selected AGNs \citep{Hainline2011} and the majority of 
type II $z\simeq 2-4$ quasars from \citet{Alexandroff2013}. This follows naturally if the metal poor 
galaxies have spectra which are deficient in the 4 Ryd photons which power He II and triply ionize oxygen.

The origin of the line emission can be clarified further by comparison to photoionization models. 
The right panel of Figure 4 shows the line ratios of RXC J2248-ID3  in the context of the AGN and stellar 
models from \citet{Feltre2016}.    RXC J2248-ID3 has log(OIII]/He II) $> 0.47$, 
which is inconsistent with line ratios expected for AGN  (log(OIII]/He II $= -3$ to $-0.5$) but 
can be easily explained by stellar models.  In particular, a hard spectrum with steep drop above 
4 Ryd is characteristic of low-metallicity stellar populations.  The precise metallicity required 
is dependent on the input stellar spectrum and may vary somewhat for single star models (like those considered 
in \citealt{Feltre2016}) and those that include binary evolution.

The presence of CIV in a two of the first few $z>6$ galaxies with deep spectra suggests that hard ionizing 
spectra may be more common in the reionization era.  However when considering whether the CIV emission in 
 RXC J2248-ID3 is typical, it is important to remember that the galaxy was selected to probe the low-mass 
regime where metallicities may indeed be systematically lower.  The existence of Ly$\alpha$ may further bias this 
selection toward low-metallicity (and  low dust content) as well.   Indeed, existing data at $z>1.5$ suggest that nebular CIV 
emitters tend to be characterized by low stellar mass  (2$\times$10$^{6}$-1.1$\times$10$^8$ M$_\odot$) 
and large equivalent width Ly$\alpha$ emission, as would be expected if the stellar populations capable 
of powering high ionization lines are only found among young, low metallicity stellar populations.  The nature of massive 
stellar populations at low metallicity remains poorly understood.  Theoretical work on massive star binary evolution 
(e.g., \citealt{Eldridge2009, deMink2014}) indicate that the lifetimes and high-energy ionizing output of 
massive stars at low metallicity may be vastly different (and higher) than classically assumed, potentially  
explaining the large luminosities now being detected in high ionization nebular lines.   

Large samples of galaxies with intercombination metal line detections
at high spectral resolution can constrain $z>6$ Ly$\alpha$ velocity offsets, a critical input 
for efforts to infer the IGM ionization state from Ly$\alpha$ emitters.  Our measurement 
of $\Delta v=235$ km s$^{-1}$  in a sub-L$^\star$ object
(M$_{\rm{UV}}=-$20.1) at $z=6.11$ falls between existing measurements of
high-luminosity objects with large Ly$\alpha$ offsets and low-luminosity
objects with small offsets at $z>6$.  In a partially neutral IGM,  a large 
velocity offset will allow line radiation to redshift further into the damping wing 
by the time it encounters intergalactic hydrogen, thereby reducing IGM attenuation relative 
to systems with smaller offsets \citep{Dijkstra2010}. This M$_{\rm{UV}}$-$\Delta v$ relationship (Figure 3b) 
will thus help create a luminosity-dependent Ly$\alpha$ fraction, consistent with emerging 
measurements (Stark et al. 2017).   This issue should be
further clarified through increasing samples expected in the near future.

To summarize, the detections of CIV and OIII] in a $z > 6$ galaxy, possibly hints at a markedly different 
underlying stellar population in typical galaxies at $z>6$ relative to those studied at lower redshift.
The detection of high ionization UV features in RXC J2248-ID3 
likely suggests that they are more common in the reionization era than previously expected.  
Taken together, this implies that extrapolations from lower redshifts may be missing a significant 
and qualitative change in the nature of of photon production in the epoch of reionization.
 
\acknowledgments
DPS acknowledges support from the 
National Science Foundation  through the grant AST-1410155.   We are grateful to Dawn Erb for providing 
Ly$\alpha$ velocity offset data from Erb et al. (2014). 

\bibliographystyle{aasjournal.bst}

\end{document}